\documentclass[aps,twocolumn,pra,tightenlines,floatfix,superscriptaddress]{revtex4-2}

\usepackage[breaklinks=true]{hyperref}
\usepackage{graphicx}
\usepackage[english]{babel}
\usepackage{amsmath}
\usepackage{amssymb}
\usepackage{times}

\begin{document}

\title{Flat band effects on the ground-state BCS-BEC crossover in atomic Fermi gases in a quasi-two-dimensional Lieb lattice}

 \author{Hao Deng}
 \affiliation{Hefei National Research Center for Physical Sciences at the Microscale and School of Physical Sciences, University 
   of Science and Technology of China, Hefei, Anhui 230026, China}
 \affiliation{Shanghai Research Center for Quantum Science and CAS Center for Excellence in Quantum Information and Quantum Physics, 
 University of Science and Technology of China, Shanghai 201315, China}
 \affiliation{Hefei National Laboratory, University of
  Science and Technology of China, Hefei 230088, China}
 \author{Chuping Li}
 \affiliation{Hefei National Research Center for Physical Sciences at the Microscale and School of Physical Sciences, University 
   of Science and Technology of China, Hefei, Anhui 230026, China}
 \affiliation{Shanghai Research Center for Quantum Science and CAS Center for Excellence in Quantum Information and Quantum Physics, 
 University of Science and Technology of China, Shanghai 201315, China}
 \affiliation{Hefei National Laboratory, University of
  Science and Technology of China, Hefei 230088, China}
 \author{Yuxuan Wu}
 \affiliation{Hefei National Research Center for Physical Sciences at the Microscale and School of Physical Sciences, University 
   of Science and Technology of China, Hefei, Anhui 230026, China}
 \affiliation{Shanghai Research Center for Quantum Science and CAS Center for Excellence in Quantum Information and Quantum Physics, 
 University of Science and Technology of China, Shanghai 201315, China}
 \affiliation{Hefei National Laboratory, University of
  Science and Technology of China, Hefei 230088, China}

 \author{Lin Sun}
 \email[Corresponding author: ]{lsun22@ustc.edu.cn}
 \affiliation{Hefei National Laboratory, University of
  Science and Technology of China, Hefei 230088, China}
 \affiliation{Shanghai Research Center for Quantum Science and CAS Center for Excellence in Quantum Information and Quantum Physics, 
 University of Science and Technology of China, Shanghai 201315, China}

 \author{Qijin Chen}
 \email[Corresponding author: ]{qjc@ustc.edu.cn}
 \affiliation{Hefei National Research Center for Physical Sciences at the Microscale and School of Physical Sciences, University 
   of Science and Technology of China, Hefei, Anhui 230026, China}
 \affiliation{Shanghai Research Center for Quantum Science and CAS Center for Excellence in Quantum Information and Quantum Physics, 
 University of Science and Technology of China, Shanghai 201315, China}
 \affiliation{Hefei National Laboratory, University of
  Science and Technology of China, Hefei 230088, China}

\date{\today}

\begin{abstract}
  The ground-state superfluid behavior of ultracold atomic Fermi gases
  with a short-range attractive interaction in a quasi-two-dimensional
  Lieb lattice is studied using BCS mean-field theory, within the
  context of BCS-BEC crossover.  We find that the flat band leads to
  nontrivial exotic effects.  As the Fermi level enters the flat band,
  both the pairing gap and the in-plane superfluid density exhibit an
  unusual power law as a function of interaction, with strongly
  enhanced quantum geometric effects, in addition to a dramatic
  increase of compressibility as the interaction approaches the BCS
  limit. As the Fermi level crosses the van Hove singularities, the
  character of pairing changes from particle-like to hole-like or vice
  versa. We present the computed phase diagram, in which a pair
  density wave state emerges at high densities with relatively strong
  interaction strength.
\end{abstract}

\maketitle

\section{Introduction}

Ultracold atomic Fermi gases in optical lattices have become an ideal
platform for quantum simulation and quantum engineering and thus have
enormous potential for exploring difficult condensed matter problems
and new quantum physics
\cite{chen2005PR,bloch2008RMP,Zwerger2012,hart2015observation,bloch2008quantum},
due to their multiple adjustable parameters, including interaction
strength, lattice depth, dimensionality, population imbalance, and
lattice geometry, etc
\cite{demarco1999onset,bartenstein2004crossover,kinast2005,zwierlein2006fermionic,partridge2006pairing}.
In particular, crossover from a BCS type of superfluidity to
Bose–Einstein condensation (BEC) of fermion pairs in an attractive
Hubbard model can be realized in an optical lattice using atomic Fermi
gases in an optical lattice \cite{jaksch1998cold}. Such a BCS-BEC
crossover has been realized by tuning the effective interaction
strength through a Feshbach resonance in trapped atomic Fermi gases
\cite{chin2010feshbach}. Such studies can help to elucidate
the underlying physics of the widespread pseudogap phenomena in
cuprate superconductors \cite{Timusk1999RPP}, which is of central
importance in understanding the mechanism of high-temperature
superconductivity \cite{chen1998PRL}.

Recently, models with a flat band have attracted great interest,
because of the associated high density of states (DOS)  and  possible quantum
geometric effect associated with  multiband of such systems.
It has been reported that
flat band and quantum geometric effect may enhance the superconducting
transition temperature with an on-site attractive interaction and may
give rise to quantum Hall states with a nonzero Chern number
\cite{cao2018N,Kopnin2011PRB,Kauppila2016PRB,Wang2011PRL,beugeling2012topological,Wang2020PRB}.
Flat bands have been studied in bipartite lattices, e.g., Lieb
lattice, magic-angle graphene moir\'e lattices, as well as perovskite,
kagome and honeycomb lattices
\cite{lieb1989two,sun2011nearly,cao2018N,tang2011high,neupert2011fractional}.

Particularly, Lieb lattices have been realized in optical lattices of
ultracold atoms \cite{taie2015coherent,schafer2020tools}.  With line
centers on a square, a Lieb lattice contains a central flat band,
which touches an upper and a lower band at the Dirac points at the
Brillouin zone corners \cite{Huhtinen2018PRB}.  There have been a
number of theoretical and experimental studies on Lieb lattices,
including lattice preparation
\cite{taie2015coherent,schafer2020tools,cui2020realization} and
associated aspects such as Chern semimetals with three spinless
fermion species \cite{palumbo2015two},  ferromagnetic and
antiferromagnetic states in a repulsive
Hubbard model at half filling
\cite{noda2009ferromagnetism,noda2015magnetism,nie2017ferromagnetic},
strain-induced superconductor-insulator transition
\cite{swain2020strain}, competition between pairing and charge density
wave at half filling using determinant quantum Monte Carlo
\cite{PhysRevB.90.094506}, as well as topologically nontrivial quantum
spin Hall effect with extra interactions
\cite{zhu2017simulating,sadeghi2022spin,beugeling2012topological,pires2022transport}.
Such rich physics warrants the investigation of the important superfluidity and
pairing phenomenon of ultracold Fermi gases in a Lieb lattice, in
order to uncover possible exotic and interesting new quantum phenomena
in the presence of a flat band.

In this Letter, we investigate the flat band effects on the
ground-state superfluid behaviors of ultracold atomic Fermi gases in a
quasi-two-dimensional Lieb lattice with a short-range attractive
interaction, with the nearest neighbor hopping only, which leads to a
zero Chern number \cite{chen2014impossibility}.  We find that the flat
band, as well as the van Hove singularities (VHS), leads to exotic
phenomena in the superfluid behavior.  When the Fermi level enters the
flat band, the fermion pairing gap $\Delta$ changes from an
exponential dependence into an unusual power law, as a function of the
interaction strength $|g|$, with an enormous compressibility $\kappa$
in the BCS limit. The superfluid density also exhibits a power law
behavior, in contrast to being a constant  in 3D free space. For a
lower number density $n$ slightly above 1 fermion per unit cell, the
fermion chemical potential $\mu$ varies \emph{nonmonotonically} with
the interaction strength in the weak interaction region, and reaches a maximum when it
crosses the VHS, signaling a change of pairing character from
particle-like at strong interactions to hole-like at weak
interactions. At the same time, $\kappa$ rises in the weak coupling
region as the Fermi level gets closer to the VHS.  In the BEC regime,
$\mu$ shows an asymptotic linear dependence
on the interaction 
$g$, leading to a density-independent $\kappa = -2/g$, with
  the pairing gap
$\Delta \sim |\mu| \sim |g|$, which is qualitatively
similar to the 3D lattice case \cite{Chien2008PRA}.  Finally, the
 phase diagram reveals that a pair density wave (PDW)
ground state emerges at intermediate pairing strength for relatively
large densities, due to strong inter-pair repulsive
interactions and relatively large pair size at intermediate pairing
strength, which is also found in dipolar Fermi gases \cite{che2016PRA}
and Fermi gases in 2D optical lattices with one continuum dimension
\cite{sun2021AdP,sun2022PRA} within the pairing fluctuation theory.

\section{Theoretical Formalism}


\begin{figure}
\centerline{\includegraphics[clip,width=3.2in]{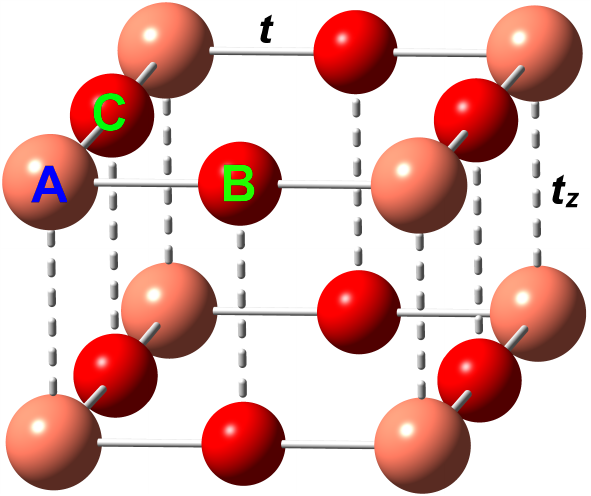}} 
\caption{
  Structure of a quasi-2D Lieb lattice. Site A represents the site in a simple square lattice, 
  while sites B and C sit at the line centers of the square in each 2D plane. 
  The solid lines represent nearest-neighbor hopping integral $t$ between site A and sites B and C. 
  The dashed lines denote the inter-layer coupling $t_z$.
  }\vspace*{-1ex}
\label{fig:Lieb}
\end{figure}

The non-interacting Hamiltonian of a Lieb lattice can be written as
\begin{equation*}
H_0=\sum_{\mathbf{k}\sigma}\hat{c}_{\mathbf{k}\sigma}^{\dag}\hat{H}_{\mathbf{k}}\hat{c}_{\mathbf{k}\sigma}
, \end{equation*}
where
$\hat{c}_{\mathbf{k}\sigma}=[C_{A\mathbf{k}\sigma},C_{B\mathbf{k}\sigma},C_{C\mathbf{k}\sigma}]^{T}$.
The subscript A denotes the site in a standard square lattice with B
and C located on the side center of the square, as shown in Fig.~\ref{fig:Lieb}. 
This leads to the Hamiltonian in
momentum space for free fermions 
\[ 
  \hat{H}_{\mathbf{k}}=\begin{bmatrix}
    d_k & a_k & b_k\\
    a_k & d_k & 0\\
    b_k & 0 & d_k
  \end{bmatrix},
\]
where $a_k=2t[1-\cos(k_x/2)]$, $b_k=2t[1-\cos(k_y/2)]$ and
$d_k=2t_z(1-\cos k_z)-\mu$ represent the hopping in the $x$ and $y$
directions, and the dispersion in the out-of-plane $\hat{z}$
direction, respectively, with $t$ and $t_z$ being the in-plane and
out-of-plane hopping integral, respectively.  We take $t_z/t=0.01$ for
the quasi-two dimensionality, set the lattice constant $a=1$, and
measure energy relative to the bottom of the lower energy band.
Diagonalizing $\hat{H}_{\mathbf{k}}$ leads to three bands with
dispersions
$\xi_{\mathbf{k}}^{\alpha}=\alpha\sqrt{2}t\sqrt{2+\cos k_x+\cos
  k_y}+2\sqrt{2}t+2t_z(1-\cos k_z)-\mu$, where $\alpha=\pm,0$ denotes
the upper, lower and flat band, respectively.  This yields in the band representation
$\displaystyle	H_0=\sum_{\mathbf{k}\alpha \sigma}\xi^{\alpha}_{\mathbf{k}}c_{\mathbf{k}\alpha \sigma}^{\dag} c_{\mathbf{k}\alpha \sigma}$,
 where $ c_{\mathbf{k}\alpha \sigma}$ is the annihilation operator in  band $\alpha$. 

The interaction Hamiltonian in the band representation is given by 
\begin{equation*}
  H_{\mathrm{int}}=\!\!\sum_{\mathbf{k}\mathbf{k}^{'}\mathbf{q}\alpha\beta}\!\! U_{\mathbf{k}\mathbf{k}^{'}\alpha\beta}c_{\mathbf{k}+\frac{\mathbf{q}}{2}\alpha\uparrow}^{\dag}
  c_{-\mathbf{k}+\frac{\mathbf{q}}{2}\alpha\downarrow}^{\dag}c_{-\mathbf{k}^{'}+\frac{\mathbf{q}}{2}\beta\downarrow}c_{\mathbf{k}^{'}+\frac{\mathbf{q}}{2}\beta\uparrow}\,,
\end{equation*}
 with band indices $\alpha,\beta=\pm,0$. The full Hamiltonian  is thus  $  H=H_0+H_{\mathrm{int}} $.
We find that with only on-site interactions, the pairing gap often varies strongly from site to site \cite{Julku2016PRL}, which necessarily leads to a large kinetic energy for the pairing field, and thus may not be energetically favorable in the superfluid state. 
 Since the nearest-neighbor hopping hybridizes different orbitals, and to be compatible with the conventional superconductivity with electron-phonon interaction induced pairing, we find it reasonable to assume a uniform short-range  intra- and inter-orbital interaction, $U_{ij}= U$, where $i,j=\{A,B,C\}$ are the orbital indices. Through a unitary transformation, this leads naturally to uniform  matrix elements of the interaction  in momentum space across all bands, with $U_{\mathbf{k}\mathbf{k}^{'}\alpha\beta}=g<0$. (See Appendix \ref{sec:AppA} for details).
This in turn gives rise to a uniform order parameter $\Delta_\alpha = \Delta$ in the mean field approximation \cite{Chamel2010PRC}.
We emphasize that, despite the short interaction range,  fermion hopping enables pairing between  sites across a large distance.


Using the BCS mean-field theory at zero temperature, the bare and full
Green's functions are given by
\begin{eqnarray*}
  G_0(K)&=&\frac{\theta(\left| k \right|-k_\text{F})}{\omega-\hat{H}_{\mathbf{k}}+i0^{+}}+\frac{\theta(k_\text{F} -\left| k \right|)}{\omega-\hat{H}_{\mathbf{k}}-i0^{+}}\,,\\
  G^{-1}(K)&=&G_0^{-1}(K)-\Sigma(K), \quad \Sigma(K)=-\Delta^2 {G}^T_0(-K)\,,
\end{eqnarray*}
respectively, 
with four momentum
$K\equiv(\omega,\mathbf{k})$, and $\theta(x)$ is the  step
function.

The number constraint $n=2\sum_{K}{\rm Tr}\,G(K)$ leads to the number
equation
\begin{equation}
  \label{eq:mu}
  n=\sum_{\mathbf{k}}\sum_{\alpha=0,\pm} \left(1-\frac{\xi_{\mathbf{k}}^{\alpha}}{E_{\mathbf{k}}^{\alpha}}\right)\,,
\end{equation}
where
$\sum_{K}\equiv \sum_{\omega}\sum_{\mathbf{k}}$, and
$E_{\mathbf{k}}^{\alpha}=\sqrt{(\xi_{\mathbf{k}}^{\alpha})^2+\Delta^2}$
is the Bogoliubov quasiparticle dispersion for band $\alpha=\pm,0$ 
with energy gap $\Delta$.  The gap equation is given by \cite{Chamel2010PRC}
\begin{equation}
  \label{eq:gap}
  0=\frac{1}{g}+\sum_{\mathbf{k}}\sum_{\alpha=0,\pm} \frac{1}{2E_{\mathbf{k}}^\alpha}\,.
\end{equation}

Equations (\ref{eq:mu}) and (\ref{eq:gap}) form a closed set of
self-consistent equations
, which can be solved for ($\mu$, $\Delta$)
as a function of $g$ in the superfluid phase.
The solution for ($\mu$, $\Delta$) in the superfluid phase should
satisfy the stability condition that the Cooper pair energy should be
nonnegative. To this end, we extract the pair dispersion using the
fluctuating pair propagator, as given in the pairing fluctuation
theory which was previously developed for the pseudogap physics in the
cuprates \cite{chen1998PRL}, and extended to address the
BCS-BEC crossover in  atomic Fermi gases \cite{chen2005PR}.
To be compatible \cite{chen2023superconductivity} with the BCS-Leggett
ground state, the pair susceptibility in the $T$-matrix
$t_\text{pg}^{-1}(Q)=1/g+\chi(Q)$ is given by
$\chi(Q)=\sum_{K}{\rm Tr}\,[G(K){G}^T_{0}(Q-K)]$, with
$Q\equiv(\Omega,\mathbf{\mathbf{q}})$, which leads to
$t_\text{pg}^{-1}(\Omega,\mathbf{q})\approx
a_0(\Omega-\Omega_\mathbf{q})$, with pair dispersion
$\Omega_{\mathbf{q}}=2B(2-\cos q_x-\cos q_y)+2B_z(1-\cos q_z)$. Here
$B$ and $B_z$ correspond to the effective pair hopping integral in the
$xy$ plane and in the $z$ direction, respectively.  The expressions
for the coefficients $a_0$, $B$ and $B_z$ can be readily derived
during the Taylor expansion.  The non-negativeness of the pair
dispersion requires that the pairing correlation length (squared)
$\xi^2 = a_0 B$ and $\xi_{z}^2 = a_0B_z$ be positive.

Superfluid density $(n_s/m)$ is an important transport property; its
dependence on the interaction  often reflects the pairing
symmetry. It takes the average of the inverse band mass in a lattice,
in contrast with the 3D continuum case where it is always given by
$n/m$ at $T=0$ in BCS theory.  Moreover, the in-plane component of
$(n_s/m)$ can be divided into a conventional and a geometric part, due
to the presence of the flat band, where the geometric term is
associated with the interband contributions, which are proportional to
a quantum metric tensor.

The expressions for the superfluid density $(n_s/m)$ are derived using
the linear response theory within the BCS framework
\cite{Scalapino1992PRL,Kosztin1998}, which is applied to the
multi-band system \cite{liang2017band}\footnote{Note that the
  $(n_s/m)$ tensor has to be calculated as a whole, not $n_s$ and $m$
  separately.}.  Similar to Ref.~\cite{liang2017band}, the in-plane
superfluid density $(n_s/m)_\parallel$ contains a conventional term
$(n_s/m)^{\rm conv}_{\parallel}$ and a geometric term
$(n_s/m)_{\parallel}^{\rm geom}$, i.e.,
\begin{equation}
  \label{eq:nsxy}
  \left(\frac{n_s}{m}\right)_{\parallel}=\left(\frac{n_s}{m}\right)^{\rm conv}_{\parallel}+\left(\frac{n_s}{m}\right)^{\rm geom}_\parallel,
\end{equation}
where
\begin{align}
  \label{eq:nsxyc}
  \left(\frac{n_s}{m}\right)^{\rm conv}_{\parallel}=&\frac{t^2}{4}\sum_{\mathbf{k}}\sum_{\alpha=\pm}\frac{\Delta^2}{{(E_{\mathbf{k}}^\alpha)}^3}\frac{\sin^2k_x+\sin^2k_y}{2+\cos k_x+\cos k_y}\,,\\
  \left( \frac{n_s}{m} \right)^{\rm geom}_{\parallel}=&\Delta^2\sum_{\mathbf{k}}\left[\left(\frac{1}{E_{\mathbf{k}}^+}-\frac{1}{E_{\mathbf{k}}}\right)\frac{\xi_{\mathbf{k}}-\xi_{\mathbf{k}}^+}{\xi_{\mathbf{k}}+\xi_{\mathbf{k}}^+}+\right.\nonumber\\
    &\left.\left(\frac{1}{E_{\mathbf{k}}^-}-\frac{1}{E_{\mathbf{k}}}\right)\frac{\xi_{\mathbf{k}}-\xi_{\mathbf{k}}^-}{\xi_{\mathbf{k}}+\xi_{\mathbf{k}}^-}\right](g_{xx}+g_{yy})\,.
  \label{eq:nsxyg}
\end{align}
Here
$g_{\mu \nu}={\rm Re}\, (\partial_\mu\langle {+}|)(1-| +
\rangle \langle +|) \partial_\nu|{+}\rangle$ is the
quantum metric tensor of the upper or  lower band, where
$ |\pm \rangle $ is the eigenvector of $\hat{H}_{\mathbf{k}}$, associated with the upper and lower  bands, respectively.
The out-of-plane component reads
\begin{equation}
  \label{eq:nsz}
  \left(\frac{n_s}{m}\right)_{z}=2t_z^2\sum_{\mathbf{k}}\sum_{\alpha=0,\pm}\frac{\Delta^2}{{(E_{\mathbf{k}}^\alpha)}^3}\sin ^2k_z\,.
\end{equation}
Note that, without using the mean-field approximated Hamiltonian,  we do not find the extra contributions to $n_s/m$ associated with the derivative of $\Delta$ with respect to the magnetic field vector potential, unlike Ref.~\cite{Huhtinen2022PRB}.

Compressibility  $\kappa$ is an important quantity in thermodynamics, which must be positive to maintain mechanical stability.
For the ground state 
\cite{guo2013compressibility}, we obtain
\begin{align}
    \kappa&=\frac{\partial n}{\partial \mu}
    =\left( \frac{\partial n}{\partial \mu} \right)_\Delta+\left( \frac{\partial n}{\partial \Delta} \right)_\mu \frac{\partial \Delta}{\partial \mu}\nonumber\\
    &=\sum_{k\alpha} \frac{\Delta^2}{(E_{\mathbf{k}}^\alpha)^3}+\frac{\left[ \sum_{k\alpha}{\xi_{\mathbf{k}}^\alpha}/{(E_{\mathbf{k}}^\alpha)^3}  \right]^2}{ \sum_{k\alpha}{1}/{(E_{\mathbf{k}}^\alpha)^3}}\,.
  \label{eq:kappa}
\end{align}
One can tell that $\kappa$ reflects the property of density of state
(DOS).  In the weak coupling regime, a larger DOS at the Fermi level
corresponds to a higher $\kappa$.  Especially, $\kappa$ reduces to the
DOS in the noninteracting limit.

The asymptotic behavior in the BEC limit can be solved analytically
with large negative $\mu \rightarrow -\infty$.
The gap and number equations yield
\begin{equation}
  \mu = \frac{(3-n)g}{2} + 2\sqrt{2}t\,,\quad
  \label{eq:BECgap}
  \Delta = \sqrt{\frac{9}{4}g^2-\mu^2}\,,
\end{equation}
respectively, with a scaling behavior qualitatively similar to
$\Delta \sim |\mu| \sim |g|$ for 3D lattice.  Thus, we obtain for all
densities the BEC asymptote 
\begin{equation}
  \label{eq:BECcmp}
  \kappa = -\frac{2}{g} .
 \end{equation}

\section{Numerical Results and Discussions}

\begin{figure}
\centerline{\includegraphics[clip,width=3.2in]{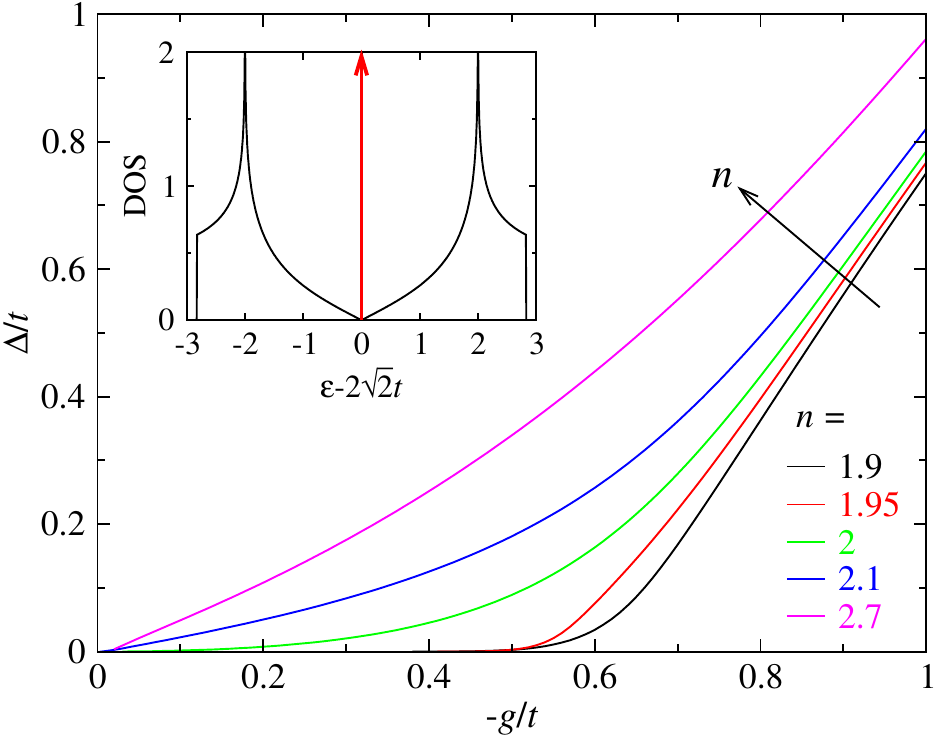}} 
\caption{$\Delta$ as a function of $-g$ for various $n$ in the weak
  coupling BCS regime, with $\mu$ close to or inside the flat band.
  Shown in the inset is the DOS for strict 2D, where the (red)
  arrow denotes the flat band.  }\vspace*{-1ex}
\label{fig:gap}
\end{figure}

\begin{figure}
\centerline{\includegraphics[clip,width=3.2in]{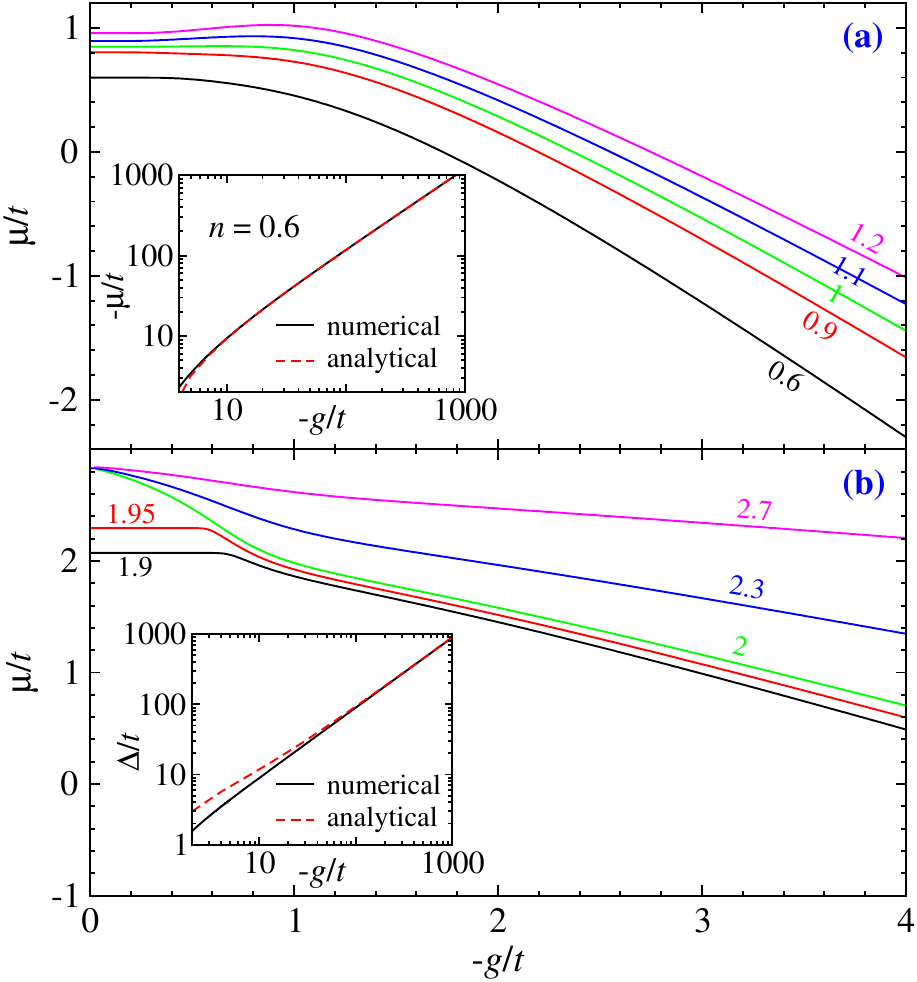}} 
\caption{Evolution of $\mu$ as a function of $-g/t$  for (a) small $0.6 \le n \le 1.2$ and (b)
  large $1.9 \le n \le 2.7$, as labeled.  In the insets,  the full numerical solution (black
  solid) for $n = 0.6$ is compared  with the analytical BEC asymptotic behavior (red dashed curves) for
  $-\mu$ and $\Delta$, respectively, as a function of $-g/t$ on a
  log-log scale.  }
\label{fig:mu}
\end{figure}

We first investigate the flat band effects on the pairing gap
behavior, when the Fermi level is close to or within the flat band.
Due to the particle-hole symmetry, we \emph{restrict} ourselves to
$n\le 3$, and the Fermi level enters the flat band for $n \ge 2$ in
the noninteracting limit.  Plotted in Fig.~\ref{fig:gap} is $\Delta$
versus $-g$ (in units of $t$) for various densities near $n\lesssim 2$
and $n\in [2,3]$ in the weak coupling  BCS regime. Shown in the inset is the DOS for
the strict 2D case, where the (red) vertical arrow denotes the flat
band, along with two VHS's in the upper and lower bands, corresponding
to $n=1$ and $5$, respectively.  For $n<2$, including $n=1.95$,
$\Delta$ exhibits an exponentially activated behavior in the weak
coupling regime, similar to that in 3D continuum and 3D cubic
lattices. This is expected in BCS theory, assuming a roughly constant
DOS near the Fermi surface.  However, as the density $n$ goes above $2$, the Fermi
level enters the flat band with the lower band fully filled, and the gap
$\Delta(g)$ as a function of the interaction exhibits an unusual power law, which can be attributed to
the breakdown of the constant DOS approximation.  Similar behavior has
also been predicted within the dynamical mean-field theory in a quasi
2D \emph{repulsive} Lieb lattice, where the magnetism as a function of the interaction
$g$ changes from exponential to power law behavior at half filling
\cite{noda2015magnetism}.  The behavior of $\Delta$ over a broad range
is shown in the inset of Fig.~\ref{fig:mu}(b) on a log-log scale for a
lower density, $n = 0.6$. Indeed, the numerical result (black solid
line) approaches its analytical BEC asymptote (red dashed line) in the
deep BEC regime, as given by Eq.~(\ref{eq:BECgap}).

This unusual power-law behavior of $\Delta$ versus $g$ can be
explained following Ref.~\cite{noda2015magnetism}, despite the
different signs of the interaction.  Using the normalized,
dimensionless notation $\overline{X} \equiv X/W$, 
where $W =4\sqrt{2}t+4t_z$ is the bandwidth, 
the rescaled, dimensionless DOS can be simplified as
$\overline{\rho}(\overline{\varepsilon}) \equiv
W\rho(\overline{\varepsilon}) =4\theta(1/2-| \overline{\varepsilon}
-1/2|)+2\delta(\overline{\varepsilon}-1/2)$, where the $\delta$
function represents the flat band.

For $n > 2$, using  $\overline{\rho}(\overline{\varepsilon})$
in the integral in the gap equation (\ref{eq:gap}),
we obtain from the gap equation (\ref{eq:gap})
\begin{equation*}   \overline{\Delta}=\frac{n-2}{8}/\mathcal{W}(\frac{n-2}{8}\exp(\frac{1}{4|\overline{g}|})),
\end{equation*}
where $\mathcal{W}$ is the Lambert $\mathcal{W}$-function, which is
the inverse function of
$f(\mathcal{W})=\mathcal{W}\exp(\mathcal{W})$ 
(See Appendix \ref{sec:AppB} for a derivation).  For
$x\gg 1$, $\mathcal{W}(x)\approx \ln(x)-\ln(\ln(x))$.  To leading
order, this yields
$\overline{\Delta}\approx \frac{n-2}{2}|\overline{g}|$ for weak
interactions. At low $n < 2$, the $\delta$-function term in
$\overline{\rho}(\overline{\varepsilon})$ becomes irrelevant, so that
we recover the ordinary exponential BCS behavior.

Next we show in Fig.~\ref{fig:mu} the behaviors of $\mu$ as a function
of $g$ for various densities $n$, so that the Fermi level changes from
(a) the lower band below ($n=0.6$) and around the VHS
($0.9 \le n \le 1.2$) to (b) near ($n=1.9, 1.95$) or inside the flat
band ($n=2,2.3, 2.7$).  As shown in the inset of Fig.~\ref{fig:mu}(a),
$-\mu$ (black solid line, for $n = 0.6$) approaches its BEC asymptotic
behavior (red dashed line) for $-g/t > 10$.  For $n\le 1$ in
Fig.~\ref{fig:mu}(a), the chemical potential $\mu$ decreases
monotonically as the pairing interaction becomes stronger, similar to
that in a regular one-band model below half filling.  However, for
$n=1.1, 1.2$, $\mu$ becomes remarkably nonmonotonic in the weak
coupling regime; $\mu$ increases first with $|g|$, and then starts to
decrease after passing a maximum.  Such nonmonotonicity is also found
in a 2D optical lattice with a strong lattice effect, which is
comprised of two lattice and one continuum dimensions
\cite{sun2021AdP,sun2022PRA}.  In a quasi-2D Lieb lattice, the lower
band has two VHS's, at $\varepsilon=(2\sqrt{2}-2)t\approx0.8284t$ and
$\varepsilon=(2\sqrt{2}-2)t+4t_z\approx0.8684t$.  For $n=1.1,1.2$,
$\mu > 0.8684t$ for small $|g|$, i.e., the Fermi level sits slightly
above the VHS's for small $|g|$, where the DOS $\rho(\varepsilon)$ has
a negative slope, so that the pairing becomes hole-like, for which
$\mu$ increases as the pairing becomes stronger, as in a 3D cubic
lattice \emph{above} half filling. Indeed, in the \emph{weak} coupling
limit, there is generally an approximate particle-hole symmetry at the
VHS, as the DOS in the vicinity of the VHS's dominates. As $|g|$
increases further, the contribution of the DOS below the VHS's comes in, so that
$\mu$ starts to decrease again.  As $n$ increases and approaches
$n=2$, the flat band gradually affects the behaviors of $\mu$ in the
BCS regime.  For $n=1.9,1.95$ in Fig.~\ref{fig:mu}(b), $\mu$ remains
nearly a constant before decreasing with $|g|$ in the BCS regime.  For
$2 \le n \le 3$, $\mu$ enters the flat band, and starts to decrease
from roughly the same noninteracting limit $\mu_0 \approx
2\sqrt{2}t$. While the nearly constant $\mu$ for $n=1.9,1.95$ is in
accord with the exponentially small gap in the BCS regime
(Fig.~\ref{fig:gap}), a power-law decrease in $\mu$ can be readily
seen for $n \in [2,3]$, commensurate with the power-law increase of
$\Delta$ as a function of $-g$.

\begin{figure}
\centerline{\includegraphics[clip,width=3.4in]{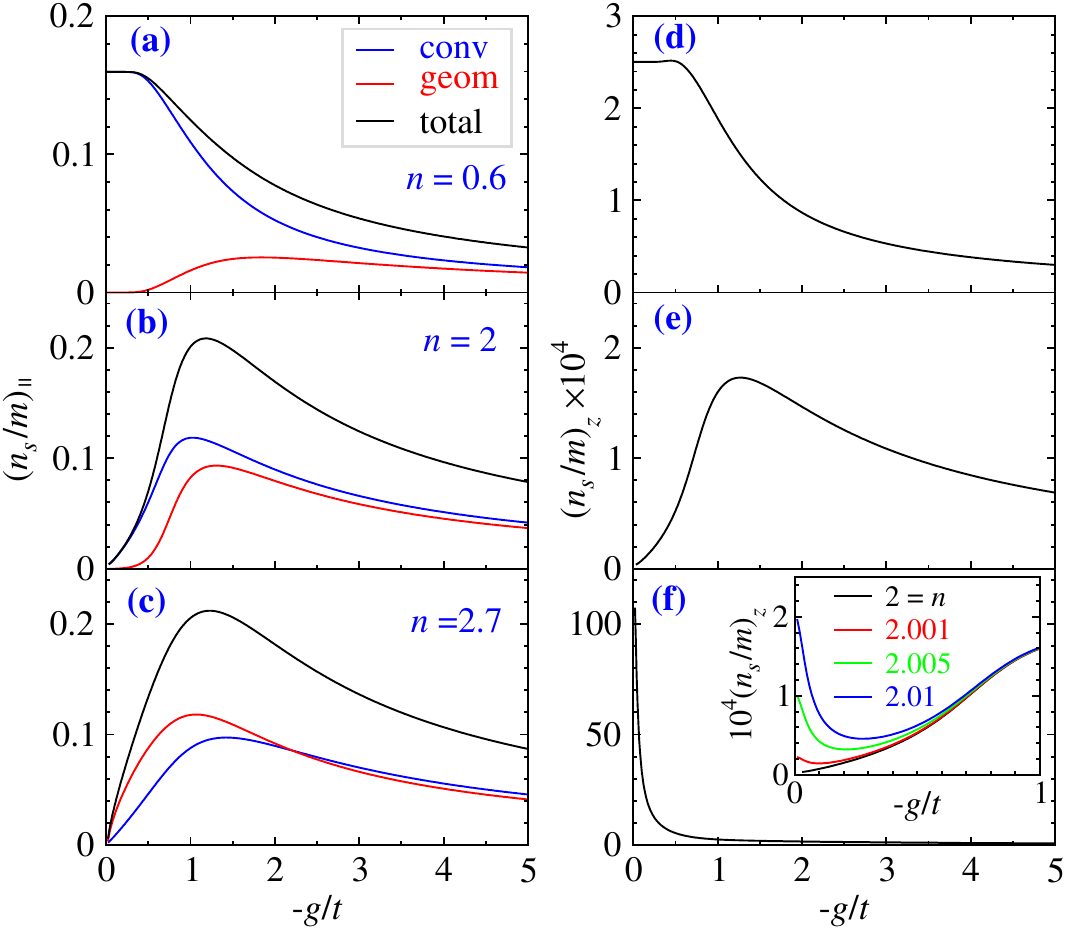}} 
\caption{Behavior of the in-plane superfluid density
  $(n_s/m)_{\parallel}$ (black curves) along with its conventional (blue
  curves) and geometric parts (red curves), as a function
  of $-g/t$, for (a) $n=0.6$, (b) $2$, (c) $2.7$.  Plotted in (d-f)
  are the corresponding out-of-plane superfluid density $(n_s/m)_{z}$.
  Shown in the inset of panel (f) is $(n_s/m)_{z}$ in the
  weak interaction regime for a series of $n\gtrsim 2$.  }
\label{fig:nSF}
\end{figure}

Shown in Fig.~\ref{fig:nSF} are the in-plane (left column) and
out-of-plane (right column) superfluid density, from top to bottom,
for $n=0.6$, $2$ and $2.7$, respectively, as well as the conventional
$(n_s/m)^{\rm conv}_{\parallel}$ (blue) and geometric part
$(n_s/m)^{\rm geom}_{\parallel}$ (red curves) of
$(n_s/m)_{\parallel}$.
For $n=0.6$, both the in-plane
$(n_s/m)_{\parallel}$ and the out-of-plane
$(n_s/m)_{z}$,  as well as the conventional
$(n_s/m)^{\rm conv}_{\parallel}$, are 
roughly constant for weak coupling, similar to the constant superfluid density $n_s/m=n/m$ in
3D continuum. On the contrary, the geometric part
$(n_s/m)^{\rm geom}_{\parallel}$ vanishes in the noninteracting limit $g\rightarrow 0$.
Commensurate with the evolution of $\Delta$ versus $-g$, as $n$
increases to $n \ge 2$, the behaviors of both
$(n_s/m)^{\rm conv}_{\parallel}$ and $(n_s/m)^{\rm geom}_{\parallel}$,
as well as $(n_s/m)_{\parallel}$, evolve into power laws as a function
of $-g$.  More importantly, $(n_s/m)^{\rm conv}_{\parallel}$ now
approaches zero as $g\rightarrow 0$, unlike in a simple cubic
lattice. This can be attributed to a few reasons. (i) Without the
out-of-plane contribution, the DOS from the upper and lower band is
zero when $\mu$ falls inside the flat band. (ii) There is a perfect
particle-hole symmetry at this $\mu$ so that the contributions from
particles and holes cancel out in $(n_s/m)_\parallel$; the pairing is
neither particle-like nor hole-like. (iii) The quantum geometric
contribution relies on a finite gap $\Delta$, and hence vanishes in
the zero $g$ limit. (iv) The flat band does not contribute to
superfluidity without the quantum geometric effect.
As $\mu$ moves away from the flat band with increasing $|g|$, the
particle-hole cancellation breaks down, leading to a rising
$(n_s/m)_{\parallel}$. As $|g|$ increases further toward the BEC
regime, $(n_s/m)_{\parallel}$ starts to decrease due to the lattice
effects. Without a geometric contribution, the out-of-plane superfluid
density $(n_s/m)_{z}$ exhibits a behavior similar to that of
$(n_s/m)^{\rm conv}_{\parallel}$ in the weak coupling regime for
$n=0.6$ and $2$. However, difference appears as $\mu$ enters the flat
band.  As shown in the inset of Fig.~\ref{fig:nSF}(f), when $n$
becomes slightly higher than 2, $(n_s/m)_{z}$ starts to increase as
$g\rightarrow 0$.  At $n= 2.7$, shown in Fig.~\ref{fig:nSF}(f), this
increase is so dramatic that $(n_s/m)_{z}$ becomes monotonically
decreasing with $|g|$.  This distinct behavior of $(n_s/m)_{z}$,
compared to $(n_s/m)^{\rm conv}_{\parallel}$, results from the
broadening of the flat band due to the small out-of-plane
dispersion.

\begin{figure}
\centerline{\includegraphics[clip,width=3.2in]{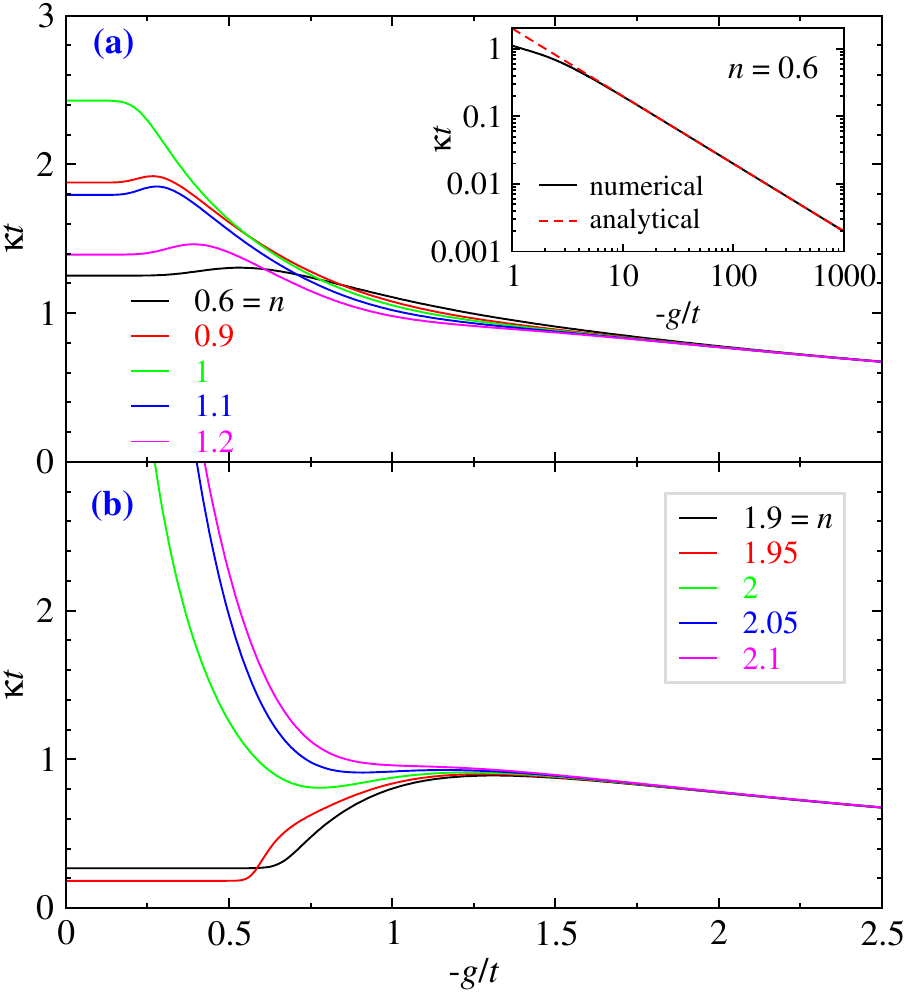}} 
\caption{$\kappa$ versus $-g$ for (a) small $n$ below ($n=0.6$) and
  close to the VHS's around $n =1 $ and (b) large $n\simeq 2$ near the
  flat band bottom.  Shown in the inset is
  the full numerical solution of $\kappa(g)$ (black solid), compared
  with its BEC asymptotic expression $\kappa = -2/g$ (red dashed
  curve) for $n = 0.6$ on a log-log scale.  }
\label{fig:cmp}
\end{figure}

Plotted in Fig.~\ref{fig:cmp} is the compressibility $\kappa$ versus
$g$ for (a) small $n$ below and around the VHS and
(b) large $n$ near and in the flat band.  Figure \ref{fig:cmp}(a)
indicates that the noninteracting value of $\kappa$ reaches a local
maximum at the VHS $n=1$ as a function of $n$, since
this value is given by the DOS. Furthermore, for $n\in [0.9, 1.2]$,
$\kappa$ varies nonmonotonically with  $g$,  largely
related to the nonmonotonic behavior of $\mu(g)$.
As $|g|$ increases into the BEC regime, all fermions pair up and the
two-body binding energy dominates $\mu$.  Consequently, $\kappa$
decreases and approaches the same $n$-independent BEC
asymptote. Indeed, as shown in the inset for $n=0.6$, $\kappa$
approaches nicely the $n$-independent analytical BEC asymptote when $-g/t > 10$.
The nearly constant $\kappa$ for $n=1.9$ and 1.95 in
Fig.~\ref{fig:cmp}(b) is clearly associated with the constant behavior
of $\mu$ in the BCS regime shown in Fig.~\ref{fig:mu}(b).  For
$n \ge 2$, $\kappa$ increases sharply as $g\rightarrow 0$. Here the
Fermi level falls within the flat band, and thus the noninteracting
$\kappa$ is given by the huge DOS. Therefore, there is a jump of
$\kappa(g=0)$ as $n$ crosses the $n=2$ boundary, as shown in
Fig.~\ref{fig:cmp}(b).

\begin{figure}
\centerline{\includegraphics[clip,width=3.in]{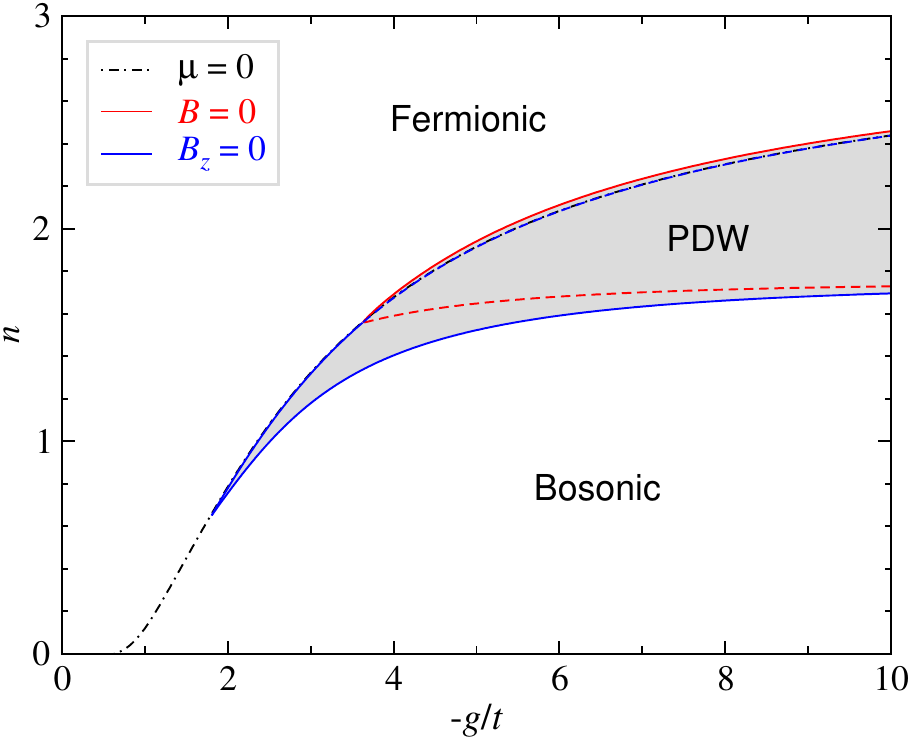}} 
\caption{Ground-state phase diagram in the $n - g$ plane.  The
  (black dot-dashed) $\mu=0$ curve divides the plane into fermionic and
  bosonic superfluid regimes.  Enclosed within the (red) $B=0$ and
  (blue) $B_z=0$ lines is a PDW ground state (gray shaded region).  }
\label{fig:phase}
\end{figure}

Finally, we present in Fig.~\ref{fig:phase} the ground-state phase
diagram in the $n$ -- $g$ plane.  Here the positivity of
$\xi^2 = a_0 B$ and $\xi_{z}^2 = a_0B_z^2$ constitutes two stability
conditions for the superfluid phase.  The (black dot-dashed) $\mu=0$
curve separates the fermionic superfluid regime on the upper left from
the bosonic superfluid regime on the lower right.  A PDW ground state
with negative $\xi^2 <0$ and/or $\xi_{z}^2<0$ emerges in the grey
shaded region, enclosed inside the (red) $B=0$ and (blue) $B_z=0$
curves. Note that the upper branch of $B=0$ overlaps nearly precisely
with $B_z=0$, and both are close to but falls slightly inside the
fermionic side of the $\mu=0$ line.  The PDW state at intermediate and
strong coupling for relatively large $n$ is associated with the strong
repulsive inter-pair interaction $U$ and relatively low kinetic energy
of the pairs, $T$, which leads naturally to Wigner
crystallization. Indeed, for $N$ pairs, the inter-pair interaction
energy $E_P$ scales as $N^2U/2$ whereas the kinetic energy $E_K$
scales as $NT$, allowing $E_P > E_K$ with a large $U$ and lattice
suppressed $T$.  Technically, the sign of $\xi^2$ (or $\xi_{z}^2$)
becomes negative between the $B=0$ (or $B_z=0$) branches, where the
pair dispersion $\tilde{\Omega}_{\textbf{q}}$ reaches a minimum at a
finite $\mathbf{q}\ne 0$, with the crystallization wave vector in the $xy$
plane (or in the $z$ direction).  The PDW state has been observed in
experiments \cite{doi:10.1126/science.abd4607}, but it is still
unclear whether it can sustain superfluidity and thus becomes a
supersolid, which will be left to a future study.  A similar PDW ground
state has been found in 3D lattices at high density
\cite{Chien2008PRA}, 2D optical lattices with strong lattice effect
\cite{sun2021AdP,sun2022PRA}, optical lattices in mixed dimensions
\cite{Zhang2017SRep}, and in dipolar Fermi gases \cite{che2016PRA}.

\section{Conclusions}

In summary, we have studied the ground-state superfluid properties of
ultracold Fermi gases in a quasi-2D Lieb lattice in the context of
BCS-BEC crossover, whose BEC asymptotic solution is derived
analytically.  We find that the flat band, together with  van Hove
singularities, have extraordinary effects on the superfluid
behavior.  When the Fermi level falls within the flat band, the
pairing gap and the in-plane superfluid density exhibit an unusual
power law in the weak coupling regime as a function of the interaction
strength $|g|$. Meanwhile, the compressibility increases sharply in the
noninteracting limit. As the chemical potential increases across the
VHS's in the lower band, the pairing becomes hole-like for weak
interactions, leading to a nonmonotonic behavior of the chemical
potential as a function of interaction strength.  A PDW ground state
emerges for intermediate and strong pairing strength with relatively
large density.  These findings for the Lieb lattice are very different
from that for pure 3D continua and 3D cubic lattices and should be
tested in future experiments.

\section{Acknowledgments}

This work was supported by the Innovation Program
for Quantum Science and Technology (Grant No. 2021ZD0301904). 

\appendix

\section{Motivation of uniform interaction across the energy bands}
\label{sec:AppA}

The attractive Hubbard model, often written in a single-band form, has been used as a formulation for superconductors in both the weak and strong coupling regimes. For a typical conventional superconductor with a pairing interaction of the electron-phonon origin, the pairing ``glue'' was mediated via the retarded ionic lattice deformation, causing electrons in the vicinity of the Fermi surface to form Cooper pairs of nearly zero center-of-mass momentum. This effective pairing interaction, originated from the Coulomb potential induced by nonzero net charge  at the deformed ion sites, necessarily propagates over a large distance, allowing for large Cooper pairs in real space. Clearly, such a Coulomb force does not distinguish between different sublattice sites in the multi-orbital situation, such as the Lieb lattice.  While the simplest one-band Hubbard model usually considers only on-site interactions, the fact that the order parameter is spatially uniform in a conventional superconductor allows one to obtain the correct result by assuming local on-site pairing, even if the electron-phonon interaction is in fact effective across a  distance.

However, such a picture may breakdown for a multi-orbital model, if one considers on-site attraction only. Indeed, this would lead to orbital selective pairing interaction, which may not necessarily be equal to each other. Therefore, the order parameter necessarily varies from site to site. An example can be seen from Ref.~\cite{Julku2016PRL}. This implies a high kinetic energy for the pairing field $\Psi$, associated with the $|\nabla\Psi|^2$ term in the Ginzburg-Landau free energy. This is clearly not energetically favorable for forming a uniform superfluid. It suggests that for a conventional superconductor, one needs to consider inter-orbital pairing interaction as well. Given the large Cooper pair size, there is essentially no real-space \emph{local} on-site pairing. Thus to make the multi-orbital Hubbard model work for weak attractions, it is desirable to assume a uniform inter-orbital and on-site interaction. (Here the ``on-site'' should be replaced by ``intra-orbital'').
Therefore, we argue that it is appropriate to use a uniform pairing interaction that does not distinguish between intra- and inter-orbitals. This in return leads to a uniform pairing interaction across all energy bands. This can be readily seen as the band representation and the orbital representation are related via a simple unitary basis transformation.

We also note that with the nearest neighbor approximation, fermion hopping occurs between different sublattice sites. Therefore, the orbital index is not a good quantum number. It is in the band index that the (bare) Hamiltonian is diagonalized. It is thus not ideal to define the order parameter in terms of the orbitals. The BCS pairing couples fermions of (nearly) opposite momenta near the Fermi level, irrespective their orbital origin. This necessarily mixes different orbitals, as the three band dispersions clearly differ dramatically from those of each orbital alone.

The fermion creation/annihilation operators are in the band and orbital representations are related via a unitary transformation, 
$C_{\mathbf{k}i\sigma}=\sum_{\alpha}S_{i\alpha}(\mathbf{k})C_{\mathbf{k}\alpha\sigma}$,
where ${i}=A,B,C$ refers to the orbitals, and $\alpha = \pm, 0$ is the band index. For the simple Lieb lattice, the transform matrix $S(\mathbf{k}) = S(\mathbf{-k})= S^\dag(\mathbf{k})$ is real, satisfying $S^\dag(\mathbf{k}) S(\mathbf{k}) = 1$.

With a short-range interaction that does not distinguish between inter- and intra-orbitals, namely $U_{ij} = U_{ii}=U$, we readily write down the two-body interaction matrix element  in the momentum space  in the center-of-mass reference frame, in terms of the orbitals,
\begin{equation*}
	\begin{split}
          H^\text{int}_{\mathbf{k}\mathbf{k}'} =&U\sum_{ij}C_{\mathbf{k}i\uparrow}^\dag C_{-\mathbf{k}i\downarrow}^\dag C_{-\mathbf{k}'j\downarrow}C_{\mathbf{k}'j\uparrow}\\
          =&U\Big[\sum_{\alpha_3 \alpha_4 i}S_{i\alpha_4}(\mathbf{k})S_{i\alpha_3}(\mathbf{k})C_{\mathbf{k}\alpha_4\uparrow}^\dag C_{-\mathbf{k}\alpha_3\downarrow}^\dag\Big]\\
          &{}\times\Big[\sum_{\alpha_1 \alpha_2 j}S_{j\alpha_1}(\mathbf{k}^{\prime})S_{j\alpha_2}(\mathbf{k}^{\prime})C_{-\mathbf{k}^{\prime}\alpha_1\uparrow}C_{\mathbf{k}^{\prime}\alpha_2\downarrow}\Big]\\
          =&U\Big(\sum_{\alpha_3 \alpha_4 }\delta_{\alpha_3 \alpha_4}C_{\mathbf{k}\alpha_4\uparrow}^\dag C_{-\mathbf{k}\alpha_3\downarrow}^\dag\Big)\Big(\sum_{\alpha_1 \alpha_2 }\delta_{\alpha_1 \alpha_2}C_{-\mathbf{k}'\alpha_1\uparrow}C_{\mathbf{k}'\alpha_2\downarrow}\Big)\\
          =&U\sum_{\alpha
            \beta}C_{\mathbf{k}\beta\uparrow}^\dag C_{-\mathbf{k}\beta\downarrow}^\dag C_{-\mathbf{k}'\alpha\downarrow}C_{\mathbf{k}'\alpha\uparrow},
	\end{split}
\end{equation*}
which has been transformed into the band representation in the last line with a uniform interaction across all bands. Here $\mathbf{k}'$ and $\mathbf{k}$ are the relative momentum (divided by 2) for the incoming and outgoing scattering fermions, respectively.

Now with the uniform  pairing interaction across bands,
$U_{\mathbf{k}\mathbf{k}^{'}\alpha\beta}=g<0$, and defining in the band representation the order parameter
\begin{align*}
  \Delta^{\alpha}_{\mathbf{k}}&=-\sum_{\beta\mathbf{k'}}U_{\mathbf{k}\mathbf{k}^{'}\alpha\beta} \left\langle
  c_{-\mathbf{k}\beta\downarrow}c_{\mathbf{k}\beta\uparrow}
                                \right\rangle\\
  &= -g\sum_{\beta\mathbf{k'}}\left\langle
  c_{-\mathbf{k'}\beta\downarrow}c_{\mathbf{k'}\beta\uparrow}
  \right\rangle = \Delta,
\end{align*}
one readily arrives at a uniform order parameter, $\Delta^{\alpha}_{\mathbf{k}}=  \Delta$,  independent of the band index. A uniform order parameter across all bands has also been considered in Ref.~\cite{Chamel2010PRC}.

In fact, in the weak coupling regime, only the gap near the Fermi level matters quantitatively. For the Lieb lattice, the Fermi level crosses only one of the three bands. In the opposite strong pairing regime, two-body physics becomes dominant so that the gap becomes quantitatively less important. This further justifies the adoption of a band-uniform pairing interaction, as a leading order approximation. 

For the electron gas in a solid, there is \emph{no natural way to create an on-site attractive interaction}, unlike the Coulomb repulsion in a repulsive Hubbard model. For the electron-phonon interaction, one may conceive that the attraction between neighboring sites is the strongest. This suggests that one may use a short-range interaction as an approximation. While the optical lattice with cold atoms may see more flexibility, one eventually needs to find a way to simulate the real solid as well.

While one certainly can consider tunable inter- and intra-band interactions (possibly via tuning the inter-orbital and intra-orbital interaction), it is reasonable to consider a uniform interaction as a first step. Indeed, in other contexts, effects of tunable inter- and intra-band interactions have been studied for multiband systems in the literature \cite{Takahashi2014,LITAK201230,Nica2017,Huang2020}. It should be noted, however, that in a typical multiband superconductor, such as MgB$_2$ \cite{TSUDA2007126},
iron pnictides and iron selenides \cite{Fernandes2022}, the multibands are often associated with multiple topologically disconnected Fermi surface sheets that are present simultaneously. This is very different from the Lieb lattice we consider. For these multi-Fermi surface cases, band selective pairing interactions, and hence band dependent pairing gaps, are more appropriate, as observed in ARPES measurements \cite{Ding2008,Fernandes2022}.

Finally, we note that the ``orbital'' in the  Lieb lattice refers to the sublattice sites, which is different from the electronic orbital states of the same atoms, such as the  $d_{xy}$ and $d_{xz}$ orbitals of Fe in iron-based superconductors. Nevertheless, both types of ``multi-orbitals'' lead to multi-bands.

\section{Derivation of the $\overline{\Delta}(\overline{g})$ relation for $n>2$, expressed in terms of the Lambert W function}
\label{sec:AppB}

For $n>2$, we substitute the (over) simplified model of $\overline{\rho}(\overline{\varepsilon})$ into the gap equation, and get
\begin{align}
    \frac{1}{|\overline{g}|}&=\int_0^{1} d\overline{\varepsilon}\,\frac{\overline{\rho}(\overline{\varepsilon})}{2\sqrt{(\overline{\varepsilon}-\overline{\mu})^2+\overline{\Delta}^2}}\nonumber\\
    &=8\int_{-\frac{1}{2}}^{0} d\overline{\varepsilon}\,\frac{1}{2\sqrt{\overline{\varepsilon}^2+\overline{\Delta}^2}}+(n-2)\frac{1}{2\overline{\Delta}}\nonumber\\
    &= 4\sinh^{-1}\left(\frac{1}{2\overline{\Delta}}\right) +(n-2)\frac{1}{2\overline{\Delta}}\nonumber\\
    &\approx 4\ln\left(\frac{1}{\overline{\Delta}}\right) +(n-2)\frac{1}{2\overline{\Delta}} 
  \label{eq1}
  \end{align}
Here $\overline{\mu} = 1/2$ and the second term on the right hand side comes from the fermion occupation in the flat band, given by $n-2$. We have assumed a small $\overline{\Delta}$ in the weak coupling regime. Therefore,
\[
  \frac{n-2}{8}\,\text{e}^{1/4|\overline{g}|} = \frac{n-2}{8\overline{\Delta}} \text{e}^{ \frac{n-2}{8\overline{\Delta}}} .
\]
Using the definition of the Lambert function, $x=\mathcal{W}(x)\text{e}^{\mathcal{W}(x)}$, finally we 
obtain
\[
\frac{n-2}{8\overline{\Delta}} = \mathcal{W}\left(\frac{n-2}{8}\,\text{e}^{1/4|\overline{g}|}\right),
\] \vspace*{-1ex}
namely, \vspace*{-1ex}
\begin{equation*}
	\overline{\Delta}=\frac{(n-2)/8}{\mathcal{W}(\frac{n-2}{8}\exp(\frac{1}{4|\overline{g}|}))}.
\end{equation*}
For small $|x|\le 1/e$, the principal branch can be Taylor expanded as \cite{Dence2013}
\[ \mathcal{W}(x) = \sum_{n=1}^\infty \frac{(-n)^{n-1}}{n!}x^n \approx x - x^2 +\frac{3}{2}x^3\,. 
\]
Relevant to our study is the large $x$ limit, which corresponds to small $\overline{\Delta}$. 
In this case, $\mathcal{W}(x)\approx\ln(x)-\ln(\ln(x))$. Therefore, to leading order, we have $\overline{\Delta}\approx \frac{n-2}{2}|\overline{g}|$ for  $n>2$ in the weak interaction regime.

Note that for $n<2$, the 2nd term in Eq.~(\ref{eq1}) is absent, and $\overline{\mu} = n/4$ based on the simplified model of $\overline{\rho}(\overline{\varepsilon})$. Then we obtain
\[
  \overline{\Delta}\approx \sqrt{n(1-n/4)}\, \text{e}^{-1/4|\overline{g}|},
\]
which recovers the conventional exponential BCS behavior. At $n=2$, it connects smoothly with the solution of Eq.~(\ref{eq1}).


%

\end{document}